\documentclass[final,5p,times,twocolumn]{elsarticle}

\usepackage[T1]{fontenc}
\usepackage[utf8]{inputenc}

\usepackage{amssymb}
\usepackage{amsmath}
\usepackage{bm}
\usepackage{dcolumn}
\usepackage{graphicx}
\usepackage{makecell}
\usepackage{glossaries}

\usepackage[version=4]{mhchem}
\usepackage[usenames,dvipsnames,svgnames]{xcolor}

\usepackage{siunitx}
\usepackage{hyperref}

\usepackage{placeins}
\usepackage{dblfloatfix}
\usepackage{balance}

\usepackage{cuted}
\usepackage{capt-of}
\usepackage{amsmath}

\DeclareSIUnit\angstrom{\text{\AA}}
\DeclareSIUnit{\atom}{atom}
\DeclareSIUnit{\step}{step}
\DeclareSIUnit{\atomstepsecond}{\atom\step\per\second}

\hypersetup{
    pdfnewwindow = true,
    colorlinks = true,
    linkcolor = Blue,
    citecolor = Blue,
    filecolor = Blue,
    urlcolor = Blue
}

\journal{Acta Materialia}

\makeatletter
\def\ps@pprintTitle{%
  \let\@oddhead\@empty
  \let\@evenhead\@empty
  \def\@oddfoot{\hfil\thepage\hfil}%
  \let\@evenfoot\@oddfoot
}
\makeatother

\biboptions{sort&compress}

\newacronym{cbn}{$c$BN}{cubic boron nitride}
\newacronym{sm}{SM}{Supplemental Material}
\newacronym{md}{MD}{molecular dynamics}
\newacronym{mlp}{MLP}{machine-learned potential}
\newacronym{dft}{DFT}{density functional theory}
\newacronym{ase}{ASE}{Atomic Simulation Environment}
\newacronym{elf}{ELF}{electron localization function} 
\newacronym{rmse}{RMSE}{root-mean-square error}

\begin{document}


\begin{frontmatter}

\title{Transitions between bulk and interfacial fracture in diamond/$c$BN heterostructures}

\author[tau]{Wei Qiu\fnref{eq1}}

\author[xjtu]{Feiyu Zhou\fnref{eq1}}
\fntext[eq1]{These authors contributed equally to this work.}
\author[imust]{Xiaonan Wang}

\author[CAEP]{Feng Xie}
\author[xjtu]{Yan Chen}

\author[xjtu]{Shengying Yue}

\author[xjtu]{Yilun Liu\corref{cor1}}
\ead{yilunliu@mail.xjtu.edu.cn}
\cortext[cor1]{Corresponding author.}

\author[xjtu]{Penghua Ying\corref{cor2}}
\ead{penghua@xjtu.edu.cn}
\cortext[cor2]{Corresponding author.}


\affiliation[tau]{
    organization={
        Department of Physical Chemistry,
        School of Chemistry,
        The Raymond and Beverly Sackler Faculty of Exact Sciences,
        and The Sackler Center for Computational Molecular and
        Materials Science,
        Tel Aviv University
    },
    city={Tel Aviv},
    postcode={6997801},
    country={Israel}
}

\affiliation[xjtu]{
    organization={
        Laboratory for Multiscale Mechanics and Medical Science,
        SV LAB,
        School of Aerospace,
        Xi'an Jiaotong University
    },
    city={Xi'an},
    state={Shaanxi},
    postcode={710049},
    country={China}
}

\affiliation[imust]{
    organization={
         School of Science, Inner Mongolia University of Science and Technology
    },
    city={Baotou},
    postcode={014010},
    country={China}
}

\affiliation[CAEP]{
    organization={
         Institute of Electronic Engineering, China Academy of Engineering Physics
    },
    city={Mianyang},
    postcode={621999},
    country={China}
}


\begin{abstract}
Whether an initially crack-free heterostructure fails at its interface or within an adjoining phase is controlled by the relative cohesion of competing atomic planes, but how interfacial chemistry, crystallographic orientation, and intermixing reshape this competition remains unclear. Here, we combine density functional theory (DFT) with a fine-tuned atomistic foundation model to resolve tensile fracture in coherent diamond/cubic boron nitride ($c$BN) heterostructures. The resulting potential reproduces independent DFT tensile responses, including an unseen (001) interface orientation. Interfacial termination, orientation, and diffusion-induced intermixing jointly determine fracture resistance and fracture-plane selection. C–N-bonded (111) and C–B-bonded (001) remain interface-controlled throughout the investigated intermixing range, whereas pristine C–B-bonded (111) fractures at a neighboring B–N plane inside $c$BN because the interface is more strongly bound. Increasing the diffusion fraction from 0 to 50.0\% causes a nonlinear decrease in fracture strength from 50.3 to 15.8 GPa and drives a bulk-to-interface transition through three regimes: $c$BN fracture up to 4.86\%, configuration-dependent competition between 7.6 and 10.1\%, and interfacial fracture at 12.5\% and above. Atom-resolved stress fields and DFT separation energetics show that this transition is associated with stress relocation and a reversal in the relative cohesion of competing planes, while electron localization analysis connects the cohesion hierarchy to termination- and orientation-dependent bonding. These results establish an atomistic framework for controlling fracture resistance and fracture pathways in strongly bonded heterostructures.

\end{abstract}



\end{frontmatter}

\flushbottom


\section{Introduction}
\label{sec:introduction}

Diamond and \gls{cbn} are two archetypal superhard solids and ultrawide-bandgap semiconductors, combining strong three-dimensional bonding networks with high elastic stiffness and thermal conductivity. Their closely matched crystal structures and complementary electronic and thermal properties make diamond/\gls{cbn} heterostructures a promising platform for high-power electronics and interfacial thermal management~\cite{qi2021measuring,wang2026interface,milne2025interface}. The performance and reliability of such heterostructures depend not only on electronic and thermal transport across the interface but also on its ability to maintain mechanical integrity and transfer load between the adjoining phases. Interfaces, defects, and crystallographic orientation can fundamentally alter how these strongly bonded lattices accommodate load, as demonstrated by nanotwinning-induced strengthening in diamond and \gls{cbn} and by size-, orientation-, and defect-dependent deformation in diamond~\cite{tian2013ultrahard,huang2014nanotwinned,nie2019diamond,dang2021elasticity,nie2020plasticity}. These advances establish the importance of microstructure and crystallography, but they do not determine whether tensile separation should occur at the interface or within one of the adjoining phases.

At the continuum scale, a crack encountering an interface between dissimilar materials may arrest, deflect along the interface, or penetrate into the neighboring material. Classical criteria describe this competition in terms of elastic mismatch and the fracture resistance of the interface relative to that of the adjoining phases~\cite{he1989crack}. These criteria, however, address the propagation of a pre-existing crack and do not resolve fracture initiation in an atomically coherent, initially crack-free heterostructure under normal tension. In this case, the competition is local: the first bond-breaking plane may lie at the interface or within a neighboring crystal plane, depending on their relative cohesion.

The experimentally reported lattice mismatch between diamond and \gls{cbn} is approximately 1.4\%, which is small enough to permit epitaxial interfaces joined by strong covalent bonds~\cite{chen2015misfit,qi2021measuring}. Experimentally realized interfaces span both coherent and defect-accommodated configurations. A coherently bonded diamond/\gls{cbn} interface has been characterized through interface-resolved phonon measurements~\cite{qi2021measuring}. The predicted interfacial thermal conductance of such coherent interfaces, on the order of 10~GW~m$^{-2}$~K$^{-1}$~\cite{wang2026interface}, further highlights the strong vibrational coupling attainable at the ideal boundary. In nanotwinned C$_2$--BN composites, diamond and \gls{cbn} domains were found to be stitched by approximately 1--2~nm-thick B--C--N interfacial regions containing \textit{sp}$^3$-hybridized C--B and C--N bonds, with interfacial dislocations accommodating the lattice mismatch~\cite{liu2016ultrahard}. Atomic-resolution microscopy likewise identified an epitaxial diamond(111)/\gls{cbn}(111) interface with direct C--B bonding and a network of Shockley partial dislocations and stacking faults~\cite{chen2015misfit}. Heterogeneous diamond--\gls{cbn} composites have achieved a fracture toughness of $16.9\pm0.8$~MPa~m$^{1/2}$ without sacrificing hardness~\cite{li2022heterogeneous}, demonstrating the mechanical potential of this heterogeneous material pair but not identifying the intrinsic separation plane of an individual interface. Because misfit dislocations and stacking faults introduce additional stress fields and potential failure sites, an ideal coherent interface provides a controlled reference state for isolating the intrinsic competition between cross-interface bonds and neighboring crystal planes. Accordingly, the present work focuses on coherent interfaces to determine how termination, orientation, and atomic intermixing govern intrinsic tensile separation and fracture-plane selection.

Electronic-structure calculations have established that C--B and C--N terminations differ in formation energy, charge transfer, and bonding, with their relative stability depending on crystallographic orientation and local stoichiometry~\cite{pickett1988superlattices,he2001polar,zhao2019electronic,milne2025interface}. Existing mechanical studies have further demonstrated that diamond/\gls{cbn} interfaces actively govern deformation under shear, indentation, and dislocation-mediated loading~\cite{yang2018nanotwinned,huang2021heterointerface,wei2025heterointerfacial}, but do not determine where separation initiates under tension normal to the interface. More broadly, cohesive failure in an adjoining phase has been observed when a strong interface efficiently transfers load~\cite{zhang2022mgti}, while termination-specific C--B and C--N bonding can select the bond-breaking site in related C/BN heterostructures~\cite{ding2016graphenehbn}. More generally, studies across composites, laminates, coatings, and heterogeneous interfaces have demonstrated that interfacial cohesion, constituent properties, residual stress, and crystallographic architecture jointly govern load transfer, fracture resistance, and crack-path selection~\cite{liu2014multimodal,wu2019interfacial,yu2017size,wang2020multiscale,babout2004competition,yan2021interlamellar,lenchuk2017cohesive}. Recent work has further shown that fracture behavior can be actively programmed through structural design and elastic instabilities~\cite{wang2026programming}; in coherent heterostructures, interfacial bonding, crystallographic orientation, and atomic intermixing may provide an atomistic route to analogous control over the fracture path. These findings motivate a local competition between the cohesion of the interface and that of neighboring crystal planes. Diffusion-induced intermixing may further reorder this competition: although chemical mixing can enhance load transfer in some interfaces~\cite{zhang2024stronginterfaces}, in diamond/\gls{cbn} it replaces an abrupt C--B or C--N plane with a chemically disordered $sp^3$-bonded region. The experimentally observed interfacial mixing can arise through atomic diffusion during high-pressure/high-temperature synthesis and vapor growth~\cite{wang2015diamondcbn,liu2016ultrahard,argoitia1994heteroepitaxy,maeda1994heteroepitaxial}. Such diffusion has also been predicted to disrupt the interface phonon modes responsible for ultrahigh thermal conductance~\cite{wang2026interface}, but its consequences for tensile fracture and fracture-plane selection have not been established. It therefore remains unknown whether intermixing simply weakens a fixed fracture plane or relocates fracture between the interface and an adjacent bulk plane, and how this response depends on bonding termination and crystallographic orientation.

Resolving these questions requires atomistic tensile simulations that accurately describe the mechanical response across bulk, interfacial, intermixed, highly strained, and bond-breaking environments. Direct \gls{dft} calculations are too costly for systematic sampling of large intermixed interfaces, motivating the development of an interatomic potential capable of resolving the subtle strength differences that determine which plane fails. \Glspl{mlp}~\cite{behler2016perspective,friederich2021machine,fan2022gpumd, muser2023interatomic, mortazavi2021firstPrinciples} provide a route to extending \gls{dft}-level accuracy to the length and time scales required for large-scale \gls{md} simulations~\cite{ying2023fullerene,mortazavi2023atomistic,ying2025advances}. Recent studies have further developed fracture-specific machine-learning force fields by explicitly sampling highly strained, bond-breaking, and nonequilibrium configurations, demonstrating their ability to resolve fracture strength and crack-path selection with near-first-principles fidelity ~\cite{ying2023fullerene,  shi2024exploring, shi2024nonequilibrium, yu2024fracture, wang2025strain, yu2026moire, guo2026carbon}.  More recently, fine-tuning pretrained atomistic foundation models on relatively small, task-specific datasets has emerged as a data-efficient alternative to training bespoke potentials from scratch, substantially reducing the amount of reference \gls{dft} data required~\cite{batatia2025foundation,zhao2025nonreciprocity,radova2025finetuning,tompa2026finetuning,liang2026nep89}. We therefore fine-tune a pretrained MACE foundation model~\cite{batatia2022mace,batatia2025foundation} using task-specific \gls{dft} data spanning the relevant bulk, interfacial, intermixed, highly strained, and bond-breaking environments. The resulting potential is validated against independent \gls{dft} tensile responses, including those for the (001) interface orientation, which was absent from the fine-tuning dataset. It is then used to perform fracture \gls{md} simulations of C--B- and C--N-bonded diamond(111)/\gls{cbn}(111) interfaces over a wide range of diffusion-induced intermixing, together with the C--B-bonded diamond(001)/\gls{cbn}(001) interface.

We further show how interfacial diffusion drives a transition from bulk to interfacial fracture in the C--B-bonded (111) system, with the sampled configurations exhibiting three regimes: fracture within the adjacent \gls{cbn} phase, configuration-dependent competition, and interfacial fracture. This transition is accompanied by a nonlinear evolution of fracture strength. Whereas the C--N-bonded (111) and C--B-bonded (001) systems remain interface-controlled over the investigated diffusion range, the pristine C--B-bonded (111) system fractures at a neighboring B--N plane within \gls{cbn}, leaving the interface intact. Diffusion-induced intermixing reduces the tensile strength of this system and shifts the preferred fracture location toward the intermixed interface. By integrating fracture simulations with atom-resolved stresses, \gls{dft} binding energies, and electronic-structure analyses, we provide a mechanistic interpretation consistent with competition between the interface and neighboring crystal planes, highlighting the roles of interfacial bonding, crystallographic orientation, and diffusion in fracture-plane selection under normal tension in initially crack-free, coherent heterostructures.

\section{Computational methods}
\label{sec:methods}

\begin{figure*}[!t] 
\centering 
\includegraphics[width=1.8\columnwidth]{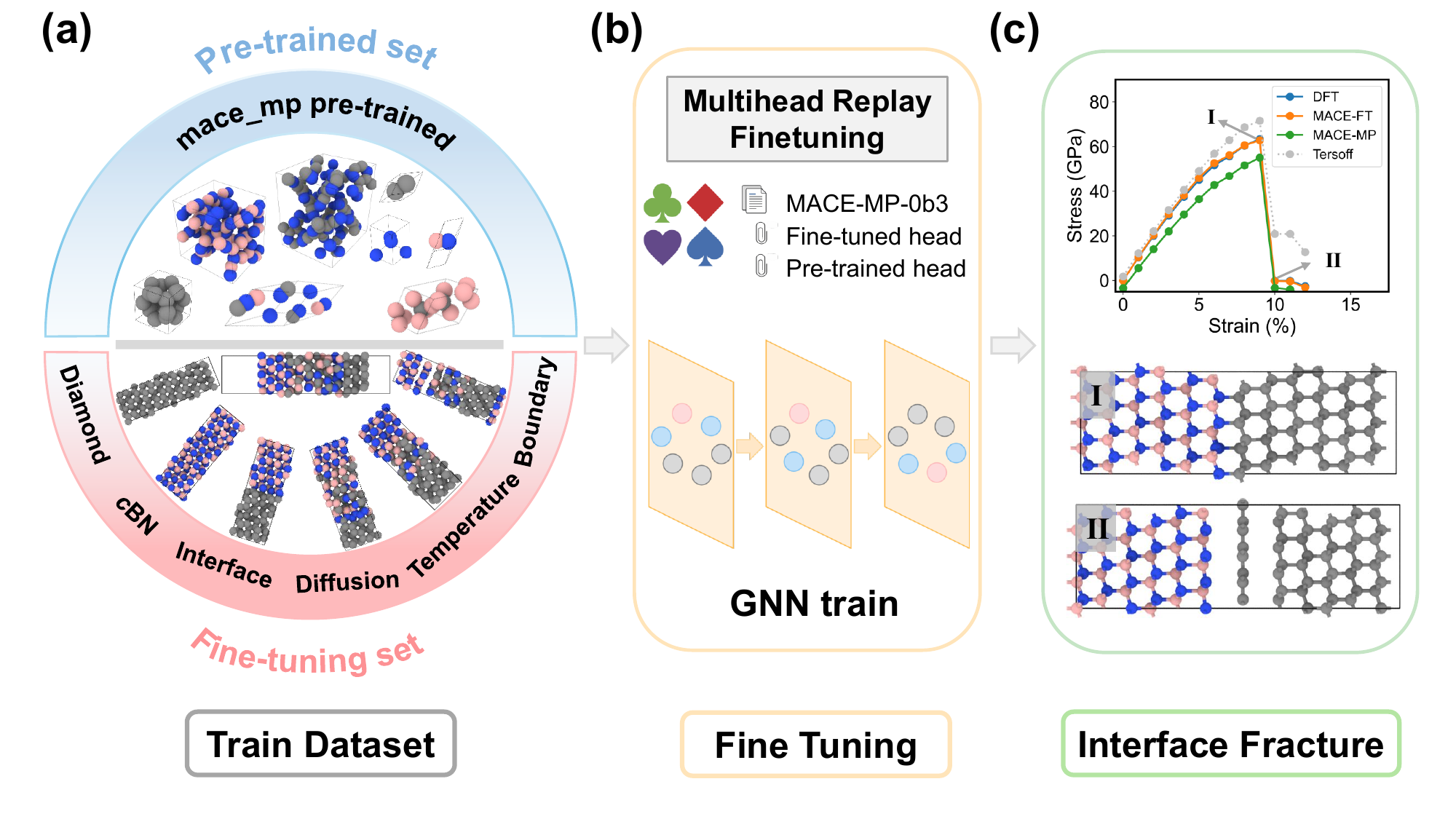}
\caption{Workflow for fine-tuning, validating, and applying a MACE foundation model  to tensile fracture of diamond/\gls{cbn} heterostructures. (a) Construction of the fine-tuning and replay datasets. The replay dataset (``Pre-trained set'') contains 200 C--B--N configurations extracted from the MACE-MP pretraining corpus. The fine-tuning dataset contains 700 DFT-labeled configurations derived exclusively from diamond (111), \gls{cbn} (111) and diamond(111)/\gls{cbn}(111) interface structures, spanning different interfacial bonding configurations, degrees of diffusion, temperatures and boundary conditions. Gray, pink and blue spheres denote C, B and N atoms, respectively. (b) Multihead replay fine-tuning of MACE-MP-0b3, in which shared representations are coupled to separate pretrained and fine-tuned output heads to retain pretrained knowledge while adapting to the target fracture environments. (c) Validation and application of the resulting MACE-FT potential. Representative tensile stress--strain responses are compared with \gls{dft} and reference interatomic potentials, with configurations I and II showing the last intact and fractured states, respectively. The validated potential is then applied to large-scale atomistic simulations of interfacial fracture.} 
\label{fig:workflow}
\end{figure*}

The overall workflow for developing and applying the machine-learned interatomic potential is summarized in \autoref{fig:workflow}. Rather than training a task-specific potential from scratch, we fine-tuned the MACE-MP-0b3-medium foundation model~\cite{batatia2022mace,batatia2025foundation}, which provides a transferable representation learned from a chemically diverse materials dataset and offers a practical balance between model capacity and computational cost. The replay dataset contains 200 configurations comprising only C, B, and N atoms extracted from the MACE-MP pretraining corpus, whereas the fine-tuning dataset contains 700 DFT-labeled configurations derived exclusively from (111)-oriented diamond, \gls{cbn}, and diamond(111)/\gls{cbn}(111) interface structures. To mitigate catastrophic forgetting during adaptation to this narrowly distributed, task-specific dataset, we employed multihead replay fine-tuning \cite{batatia2025foundation, tompa2026finetuning}, which retains relevant pretrained knowledge while specializing the model for interfacial, highly strained, and bond-breaking environments. The resulting model, denoted MACE-FT, reproduces the \gls{dft} tensile-fracture response of C--B [111] more accurately than MACE-MP and Tersoff (\autoref{fig:workflow}(c)). Its transferability was subsequently evaluated using independent \gls{dft} tensile benchmarks, including a diamond(001)/\gls{cbn}(001) interface that was absent from the fine-tuning dataset. After validation, MACE-FT was applied to systematic large-scale tensile-fracture simulations covering different interfacial bonding configurations, crystallographic orientations, and diffusion percentages.  The fracture mechanisms were subsequently examined by combining atom-resolved stress analysis with \gls{dft} separation energetics and electronic-structure calculations, providing complementary mechanical, energetic, and electronic descriptions of fracture-plane selection.  Details of each step are provided in the following subsections.

\subsection{Model setup}
\label{subsec:model}
Diamond and \gls{cbn} were constructed using cubic lattice constants of 3.5668 and 3.627~\AA, respectively. Evaluated relative to their mean lattice constant, these values yield an overall lattice mismatch of 1.67\%. The slight difference from the experimentally reported value of approximately 1.4\% arises from the specific lattice constants adopted in the present models. For the (111) models, the cell axes were oriented along $z\parallel[111]$, $x\parallel[1\bar{1}0]$, and $y\parallel[11\bar{2}]$. To construct coherent diamond(111)/\gls{cbn}(111) interfaces, the common in-plane lattice dimensions were set to the arithmetic mean of the corresponding diamond and \gls{cbn} dimensions. This construction results in approximately 0.84\% biaxial tension in diamond and 0.83\% biaxial compression in \gls{cbn}. Interfaces with either C--B or C--N interfacial bonding were considered.

The structural configurations used for fine-tuning were constructed exclusively from the (111)-oriented structural family using small cells with a $1\times1$ in-plane repeat. They included bulk diamond, bulk \gls{cbn}, pristine and intermixed diamond(111)/\gls{cbn}(111) interfaces, fully periodic cells, and non-periodic slab cells with a total vacuum thickness of 15~\AA\ along the interface-normal direction. Additional thermally perturbed, highly strained, and bond-breaking configurations were sampled from tensile trajectories at 300 and 900~K. In total, 700 configurations were selected for DFT labeling and subsequent fine-tuning. Importantly, no (001)-oriented structures were included in this dataset.

Larger heterostructures were constructed for the production fracture simulations. The standard diamond(111)/\gls{cbn}(111) fracture models contained 3456 atoms, with in-plane dimensions of approximately $15.260\times17.621$~\AA$^2$ and a solid thickness of approximately 74.760~\AA\ along $[111]$. Vacuum layers of 30~\AA\ were introduced at both ends along the interface-normal direction to suppress interactions between periodic images.

To represent interfacial intermixing induced by atomic interdiffusion, C atoms were randomly exchanged with B or N atoms within six atomic layers centered on the interface. Here, random intermixing represents the chemical mixing associated with interdiffusion, with the prescribed six-layer region serving as an idealized representation of localized interfacial mixing motivated by the nanometer-scale B–C–N interfacial regions observed experimentally~\cite{liu2016ultrahard}. The degree of intermixing was characterized by the fraction of exchanged atoms, hereafter referred to as the diffusion percentage. For the C--B-bonded and C--N-bonded diamond(111)/\gls{cbn}(111) interfaces, diffusion percentages of 0.0\%, 10.1\%, 20.1\%, 29.9\%, 39.9\%, and 50.0\% were examined. Three independently randomized structures were generated for each nonzero diffusion percentage and interfacial bonding configuration to account for the configurational variability associated with random intermixing.

To examine crystallographic-orientation effects, C--B-bonded diamond(001)/\gls{cbn}(001) heterostructures were constructed independently. Both the interface normal and tensile loading direction were along $[001]$. The standard (001) fracture models contained 2560 atoms, with in-plane dimensions of approximately $14.387\times14.387$~\AA$^2$ and a solid thickness of approximately 70~\AA\ along $[001]$. Vacuum layers of 30~\AA\ were likewise introduced at both ends along the interface-normal direction. Diffusion percentages of 0.0\%, 9.9\%, 19.8\%, 30.2\%, 40.1\%, and 50.0\% were considered. Because the fine-tuning dataset contained only (111)-oriented structures, these (001) heterostructures also served as an out-of-training-orientation test of model transferability.

\subsection{DFT calculations}
\label{subsec:dft}

\gls{dft} calculations were performed using the Vienna \textit{Ab initio} Simulation Package (VASP)~\cite{hohenberg1964inhomogeneous,kohn1965selfconsistent,kresse1996vasp}, the Perdew--Burke--Ernzerhof generalized-gradient approximation for exchange and correlation~\cite{perdew1996pbe}, and the projector augmented-wave method~\cite{blochl1994paw,kresse1999paw}. All \gls{dft} calculations used a plane-wave cutoff of 600~eV, a $\Gamma$-centered $k$-point mesh with a spacing of 0.2~\AA$^{-1}$, Gaussian smearing of 0.05~eV, and an electronic convergence criterion of $10^{-7}$~eV. Structural relaxations were considered converged when the maximum residual atomic force was below 0.01~eV~\AA$^{-1}$. \gls{dft} calculations served three purposes: generating reference labels for fine-tuning, providing independent tensile benchmarks, and characterizing the separation energetics and electronic structure associated with fracture.

For the 700 fine-tuning configurations, single-point calculations were performed using the settings described above, and the resulting energies, atomic forces, and stresses were used as \gls{dft} reference labels for fine-tuning.

Independent \gls{dft} tensile calculations were carried out to validate the fine-tuned potential outside the training configurations. Uniaxial strain was applied normal to the selected crystallographic plane in increments of 1\%. At each strain increment, the atomic coordinates and the two in-plane lattice parameters were relaxed while the imposed normal strain was held fixed. The resulting stress--strain responses were used to independently assess the accuracy and transferability of the fine-tuned potential.

To quantify the relative cohesion of competing fracture planes, \gls{dft} binding energies were calculated for bulk diamond and \gls{cbn} cleavage planes, C--B- and C--N-bonded interfaces, and the B--N plane adjacent to the C--B-bonded diamond(111)/\gls{cbn}(111) interface. For each relaxed non-periodic structure, two fragments were generated by separating the structure across the selected plane while retaining the atomic positions and the parent simulation cell. Single-point calculations were then performed for the intact structure and the two fragments. The binding energy per unit area was evaluated as
\begin{equation}
E_{\mathrm{bind}}=
\frac{E_{\mathrm{whole}}-E_{\mathrm{A}}-E_{\mathrm{B}}}{A},
\end{equation}
where $E_{\mathrm{whole}}$, $E_{\mathrm{A}}$, and $E_{\mathrm{B}}$ are the energies of the intact structure and the two separated fragments, respectively, and $A$ is the cross-sectional area of the separation plane. Under this definition, a more negative $E_{\mathrm{bind}}$ corresponds to stronger cohesion.

The bonding characteristics of representative pristine diamond/\gls{cbn} interfaces were further examined using the \gls{elf} method~\cite{becke1990simple}. Cross-sectional \gls{elf} maps passing through the relevant interfacial bonds were used to compare the C--B- and C--N-bonded diamond(111)/\gls{cbn}(111) and diamond(001)/\gls{cbn}(001) interfaces (see \autoref{fig:elf}).

\subsection{Multihead replay fine-tuning}
\label{subsec:FT}
The pretrained MACE-MP-0b3-medium model~\cite{mace_mp_0b3,batatia2025foundation}, hereafter referred to as MACE-MP, was used as the foundation model. The \textsc{MACE} package (version 0.3.15)~\cite{batatia2022mace} was used for model fine-tuning and static calculations in conjunction with the \textsc{ASE} package~\cite{larsen2017atomic}. The fine-tuning dataset comprised 700 DFT-labeled target configurations from the (111)-oriented diamond/\gls{cbn} structural family and a replay dataset of 200 representative C--B--N configurations selected by farthest-point sampling~\cite{imbalzano2018fps} from the MPtrj dataset~\cite{deng2023chgnet} used in the pretraining of MACE-MP (see \autoref{fig:workflow}(a)). Multihead replay fine-tuning was performed using shared model representations with separate output heads for the target and replay datasets. The target and replay datasets were assigned relative weights of 10 and 1, respectively. The loss weights for energy, atomic forces, and stress were set to 1, 100, and 1, respectively. Training was performed for up to 200 epochs with a batch size of 10, double-precision arithmetic, and stochastic weight averaging~\cite{izmailov2018swa}. Atomic reference energies were inherited from the foundation model. The resulting fine-tuned potential is hereafter referred to as MACE-FT. The fine-tuning and replay datasets, training inputs, and resulting MACE-FT model are publicly available in a Zenodo repository~\cite{qiu_2026_22074537}. 

\begin{figure*}[htbp]
\centering
\includegraphics[width=\textwidth]{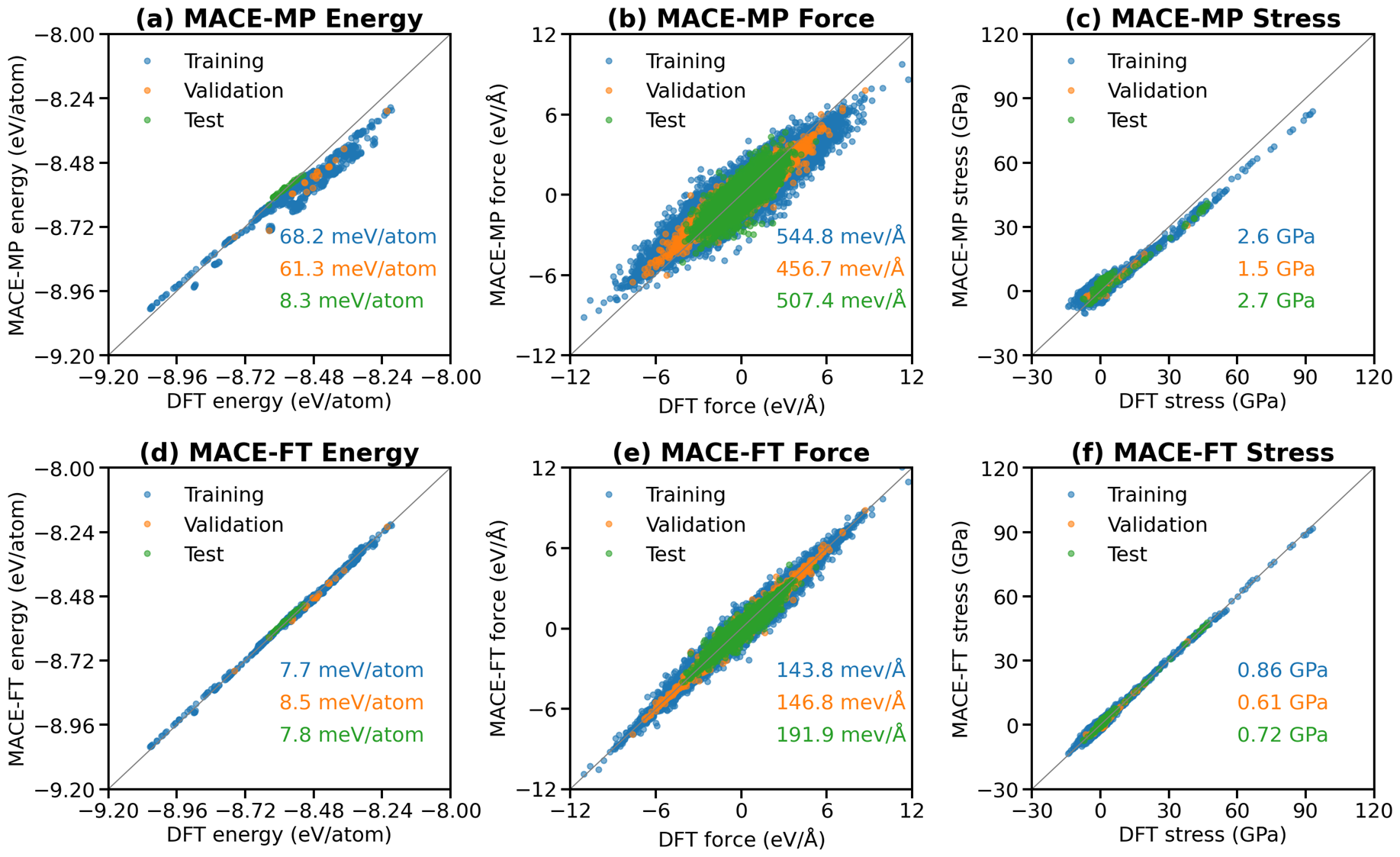}
\caption{Comparison between \gls{dft} reference values and the predictions of the pretrained MACE-MP model (a--c) and the fine-tuned MACE-FT model (d--f). Energy per atom, atomic-force components, and stress components are shown in (a,d), (b,e), and (c,f), respectively. Blue, orange, and green symbols denote the training, validation, and test subsets, respectively. The gray diagonal lines indicate perfect agreement and the colored values give the corresponding RMSEs for the three subsets.}
\label{fig:RMSE}
\end{figure*}

\subsection{Phonon dispersion calculations}
\label{subsec:phonon}
Phonon dispersions were calculated for bulk diamond, bulk \gls{cbn}, and the diamond(111)/\gls{cbn}(111) superlattice to further assess the lattice-dynamical accuracy of MACE-MP and MACE-FT. The \gls{dft} reference dispersions were taken from our previous work~\cite{wang2026interface}, where the second-order force constants were obtained using density-functional perturbation theory with the PBE functional and subsequently analyzed using the \textsc{Phonopy} package. In the present work, the MACE phonon calculations were performed using the finite-displacement approach through \textsc{Phonopy} package together with \textsc{ASE} package, with $4\times4\times4$ supercells. The dispersions were evaluated along the same high-symmetry paths as the corresponding \gls{dft} references (see \autoref{fig:dft-compare}(d)--(f)).

\subsection{Fracture simulations}
\label{subsec:MD}
Finite-temperature \gls{md} simulations of tensile fracture were performed in \textsc{LAMMPS} package~\cite{thompson2022lammps} using MACE-FT at 300~K. Periodic boundary conditions were applied in all directions, while the vacuum layers along $z$ suppressed interactions between periodic images. Atomic layers with a thickness of 5~\AA\ at both ends were treated as rigid grips. Before loading, each structure was energy-minimized and equilibrated for 10~ps using Nosé--Hoover temperature and pressure control~\cite{nose1984unified,hoover1985canonical,martyna1994constant}. Only the lateral cell dimensions were barostatted to zero pressure, while the $z$ cell length was fixed.

Uniaxial tension was applied normal to the interface by moving the two grips at equal and opposite velocities. Engineering strain rates of $3\times10^{9}$ and $3\times10^{8}$~s$^{-1}$ were considered with a time step of 1~fs. Engineering strain was calculated from the initial grip separation, and tensile stress was obtained from the mean grip reaction force divided by the initial cross-sectional area. Three replicas were performed at each finite-rate condition. Independently intermixed structures were used at nonzero diffusion percentage, whereas different initial velocity seeds were used for the pristine interface.

In addition to the finite-temperature MD simulations, quasi-static fracture simulations were performed at 0~K by displacing the grips in strain increments of 0.5\%. After each increment, the mobile atoms and lateral cell dimensions were energy-minimized while the imposed normal displacement was fixed. One pristine structure and three independently intermixed structures at each nonzero diffusion percentage were evaluated.

Atom-resolved stresses were evaluated by post-processing selected \gls{md} trajectory frames using MACE-FT together with the \gls{ase} package~\cite{larsen2017atomic}. Following the microscopic stress framework for many-body potentials~\cite{irving1950statistical,thompson2009general}, the atomic virial and local configurational stress were calculated as
\begin{equation}
\begin{aligned}
\mathbf{W}_i
&=
\sum_{j\ne i}
\mathbf{r}_{ij}\otimes
\frac{\partial U_j}{\partial\mathbf{r}_{ji}},
\\[3pt]
\boldsymbol{\sigma}_i
&=
-\frac{\mathbf{W}_i+\mathbf{W}_i^{\mathrm{T}}}
{2\Omega_i},
\end{aligned}
\label{eq:atomic-stress}
\end{equation}
where $\mathbf{r}_{ij}=\mathbf{r}_j-\mathbf{r}_i$, $U_j$ is the atomic-energy contribution associated with atom $j$, $\mathbf{W}_i$ is the atomic virial tensor, and $\Omega_i$ is the effective atomic volume. To eliminate stress dilution caused by the vacuum layer, the effective material volume was recalculated for every trajectory frame. Atomic positions were projected along the direction normal to the interface, and the largest periodic gap was identified as the vacuum region. The instantaneous material thickness was obtained by subtracting this gap from the cell height, and the material volume was calculated by multiplying this thickness by the cross-sectional area. An equal-volume partition was then adopted, with the material volume divided equally among all atoms. This frame-dependent treatment accounts for the change in material thickness during tensile deformation. The corrected atomic stresses were scaled such that their arithmetic mean recovers the corresponding vacuum-corrected macroscopic configurational stress.

The local tensile component $\sigma_{zz}$ was used to characterize load transfer and stress localization, with $z$ normal to the interface. For each interface configuration and diffusion percentage, the last intact frame immediately preceding the macroscopic stress drop was analyzed (see \autoref{fig:stressfield}). Positive $\sigma_{zz}$ denotes tension.

\section{Results and discussion}
\subsection{Validation of MACE-FT model}

\begin{figure*}[htbp]
\centering
\includegraphics[width=\textwidth]{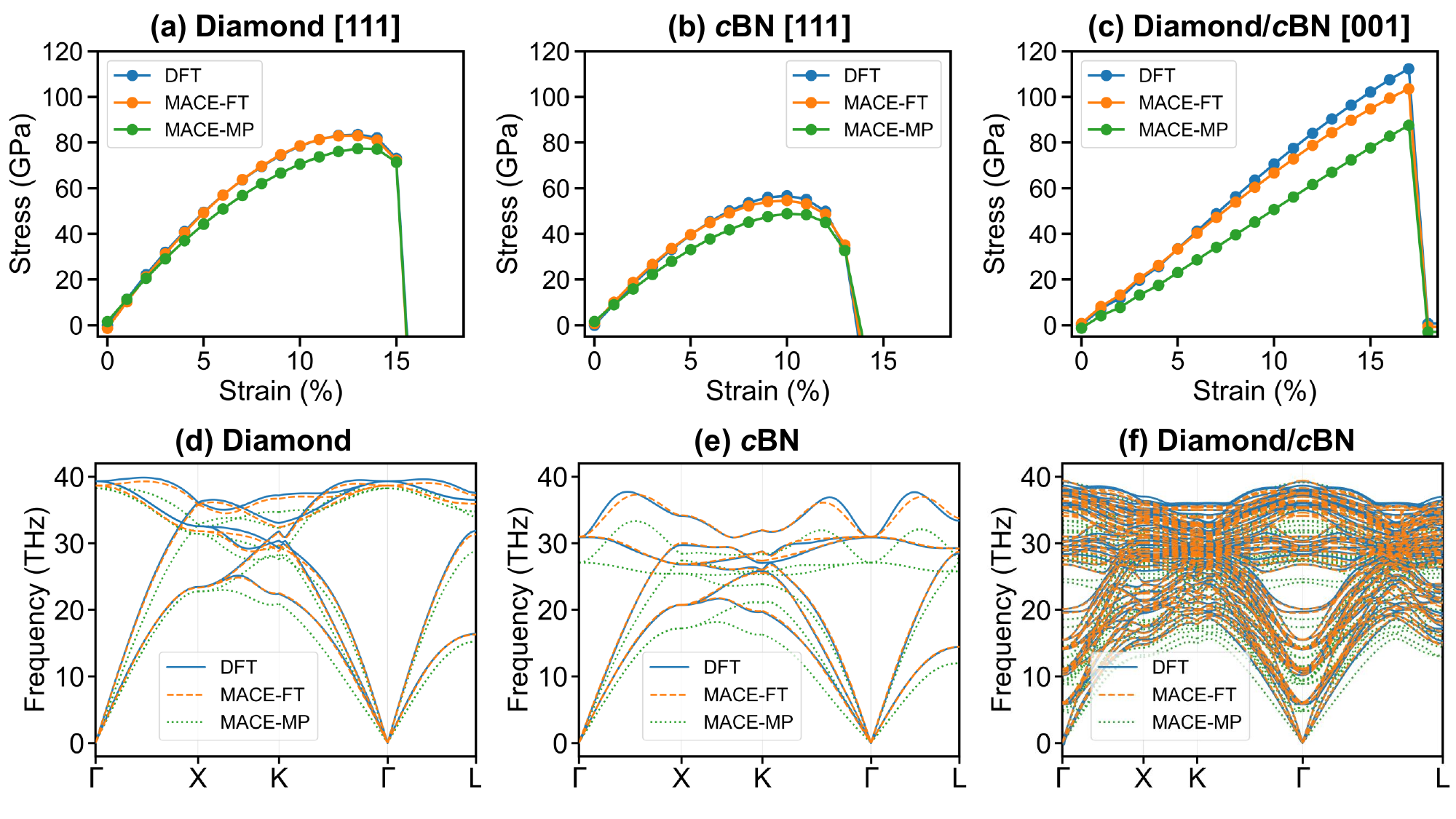}
\caption{Independent DFT benchmarks for the tensile responses of (a) bulk diamond loaded along $[111]$, (b) bulk \gls{cbn} loaded along $[111]$, and (c) the C--B-bonded diamond(001)/\gls{cbn}(001) heterostructure loaded along $[001]$. Blue, orange, and green curves represent \gls{dft}, MACE-FT, and MACE-MP results, respectively. (d)--(f) Phonon dispersions of (d) bulk diamond, (e) bulk \gls{cbn}, and (f) the diamond(111)/\gls{cbn}(111) superlattice predicted by \gls{dft}, MACE-FT, and MACE-MP. The \gls{dft} phonon reference results are taken from our previous work~\cite{wang2026interface}. Solid blue, dashed orange, and dotted green curves denote \gls{dft}, MACE-FT, and MACE-MP results, respectively.}

\label{fig:dft-compare}
\end{figure*}

Before examining fracture behavior in the large-scale simulations, we evaluated the accuracy and transferability of MACE-FT using training, validation, and test datasets together with independent tensile and phonon-dispersion benchmarks. Of the 700 target configurations, 680 and 20 were randomly selected for training and validation, respectively. The test set comprised 40 configurations selected from a MACE-FT tensile-fracture trajectory of diamond(111)/\gls{cbn}(111) with 34.4\% diffusion at 300~K. As shown in \autoref{fig:RMSE}, fine-tuning  reduces the errors in atomic forces and stresses while maintaining comparable energy accuracy. On the independent test set, the force \gls{rmse} decreases from 507.4~meV~\AA$^{-1}$ for MACE-MP to 191.9~meV~\AA$^{-1}$ for MACE-FT, and the stress \gls{rmse} decreases from 2.7 to 0.72~GPa. Meanwhile, the energy \gls{rmse} remains comparable, changing from 8.3~meV/atom for MACE-MP to 7.8~meV/atom for MACE-FT.

Beyond the held-out test set, the accuracy and transferability of MACE-FT were further evaluated using independent \gls{dft} tensile responses (\autoref{fig:dft-compare}(a)--(c)). For bulk diamond loaded along $[111]$ (\autoref{fig:dft-compare}(a)) and bulk \gls{cbn} loaded along $[111]$ (\autoref{fig:dft-compare}(b)), MACE-FT closely reproduces the \gls{dft} stress--strain responses, including the peak tensile stresses and fracture-associated stress drops. Similar agreement is obtained for the C--B-bonded diamond(001)/\gls{cbn}(001) heterostructure loaded along $[001]$ (\autoref{fig:dft-compare}(c)), whereas the pretrained MACE-MP model substantially underestimates its tensile response. Because the fine-tuning dataset contains only (111)-oriented structures, the agreement in \autoref{fig:dft-compare}(c) demonstrates transferability to a crystallographic orientation absent from the task-specific training data.

Phonon dispersions provide a complementary validation of the near-equilibrium curvature of the potential-energy surface. \autoref{fig:dft-compare}(d)--(f) compare  the MACE predictions with \gls{dft} results for bulk diamond, bulk \gls{cbn}, and the diamond(111)/\gls{cbn}(111) superlattice. The \gls{dft} reference dispersions are taken from our previous study of phonon transport across diamond/\gls{cbn} interfaces~\cite{wang2026interface}. For both bulk diamond and \gls{cbn}, MACE-FT reproduces the \gls{dft} acoustic and optical branches substantially more accurately than the pretrained MACE-MP model. More importantly, the same improvement is retained for the diamond(111)/\gls{cbn}(111) superlattice, where the dense folded phonon branches provide a stringent test of heterogeneous C--B--N environments. MACE-FT closely follows the \gls{dft} branch topology and characteristic frequencies, whereas MACE-MP exhibits noticeable frequency softening and shifts for several modes. Such phonon softening has been identified as a systematic tendency of universal \glspl{mlp}, including MACE-MP, and has been associated with underestimation of the local curvature of the potential-energy surface~\cite{deng2025systematic}. Together with the held-out test-set errors and independent tensile benchmarks, these results support the accuracy and transferability of MACE-FT for the extensive fracture simulations discussed below.

\subsection{Competition between interfacial and bulk fracture}
\label{subsec:main-fracture}

\begin{figure*}[htbp]
\centering
\includegraphics[width=\textwidth]{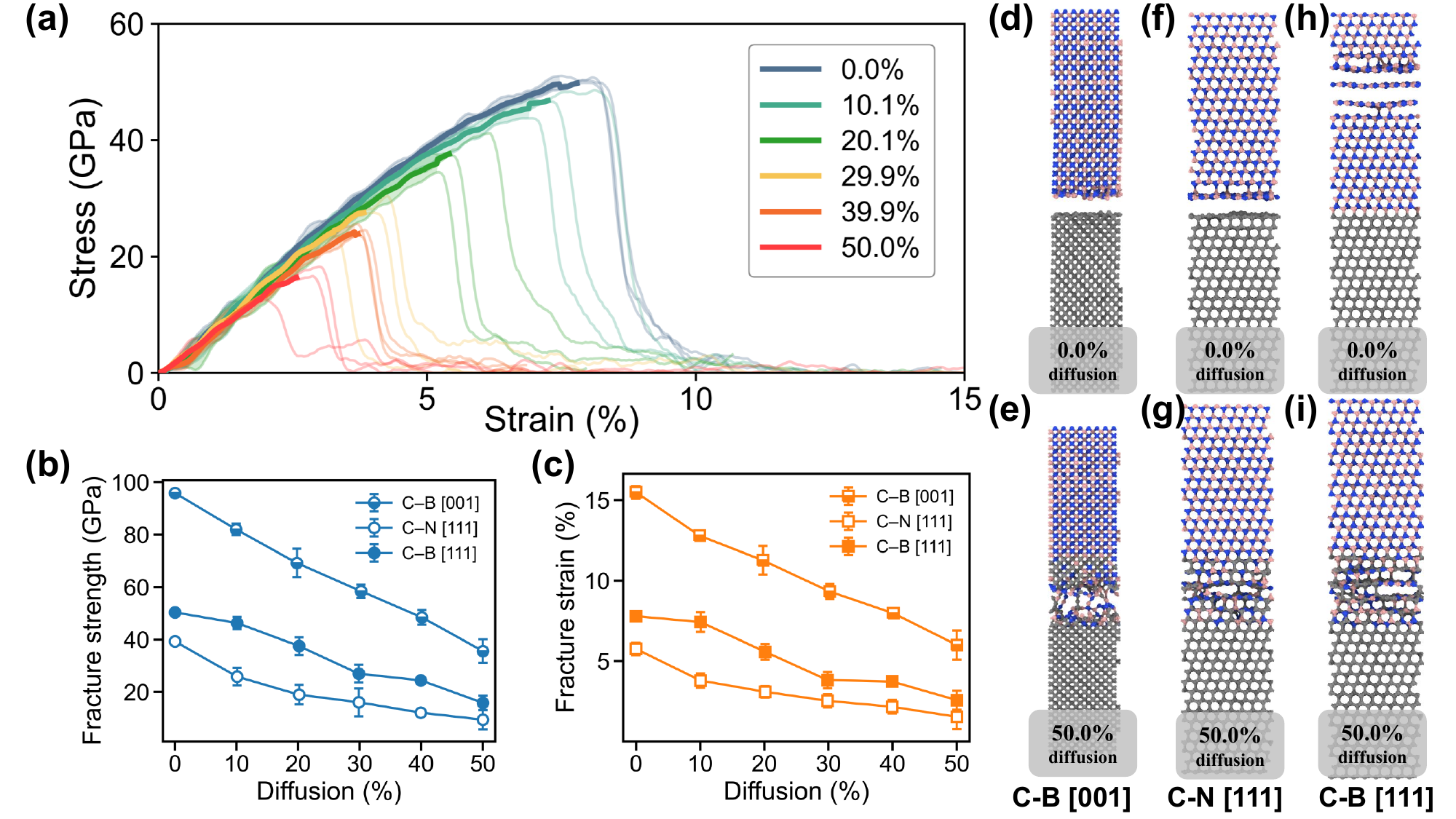}
\captionof{figure}{Effects of interfacial bonding, loading orientation, and diffusion percentage on tensile fracture in diamond/\gls{cbn} heterostructures. (a) Stress--strain responses of the C--B-bonded diamond(111)/\gls{cbn}(111) interface under $[111]$ tension at diffusion percentages of 0.0, 10.1, 20.1, 29.9, 39.9, and 50.0\%. Pale curves, dark curves, and shaded bands represent individual simulations, their mean responses, and the corresponding minimum-to-maximum ranges, respectively. (b) Fracture strength and (c) fracture strain as functions of diffusion percentage for C--B [001], C--N [111],  and C--B [111]. (d--i) Fractured configurations at 0.0 and 50.0\% diffusion for C--B [001] (d,e), C--N [111] (f,g), and C--B [111] (h,i), visualized using \textsc{OVITO} package~\cite{stukowski2010visualization}.}

\label{fig:main}
\end{figure*}

\begin{figure}[htbp]
\centering
\includegraphics[width=\columnwidth]{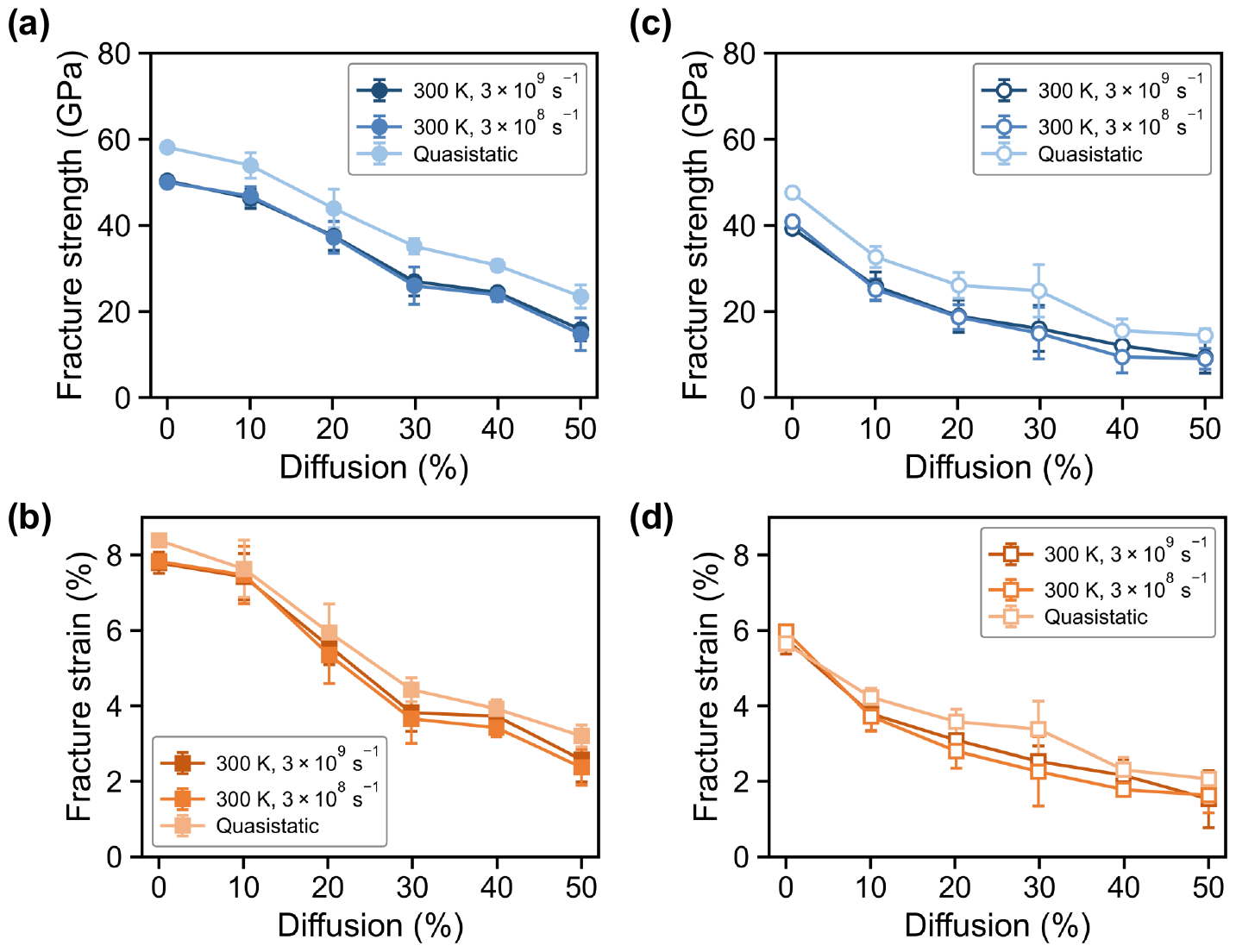}
\caption{Diffusion-dependent fracture properties of diamond(111)/\gls{cbn}(111) interfaces under different loading protocols. (a,b) Fracture strength and fracture strain of C--B [111], respectively; (c,d) corresponding results for C--N [111]. Dark, intermediate, and light colors represent 300~K molecular dynamics simulations at $3\times10^{9}$~s$^{-1}$, 300~K simulations at $3\times10^{8}$~s$^{-1}$, and quasi-static loading, respectively. Symbols and error bars denote the mean values and standard deviations obtained from independently generated structures.}
\label{fig:rate}
\end{figure}

\begin{figure}[htbp]
\centering
\includegraphics[width=\columnwidth]{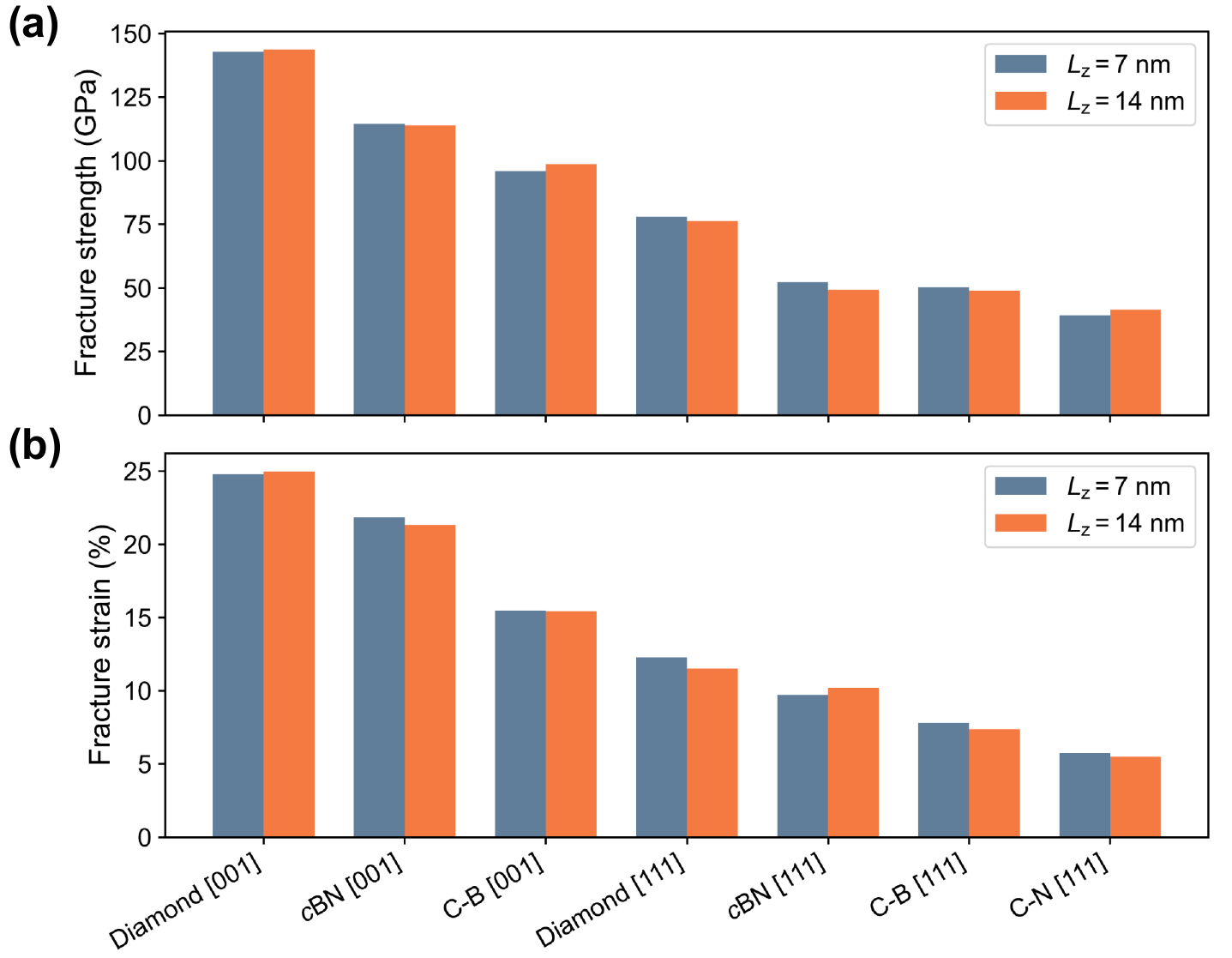}
\caption{Effect of model length on the fracture strength of ideal structures without diffusion. (a) Bulk diamond, bulk \gls{cbn}, and C--B [001] under $[001]$ tension. (b) Bulk diamond, bulk \gls{cbn}, C--B [111], and C--N [111] under $[111]$ tension. Blue and orange bars represent nominal solid lengths of approximately 7 and 14~nm along the loading direction, respectively. Error bars for the shorter C--B and C--N heterostructures denote standard deviations from three independent simulations.}
\label{fig:length}
\end{figure}

\autoref{fig:main} compares the tensile fracture behavior of three representative diamond/\gls{cbn} interfaces. Hereafter, C--B [111] and C--N [111] denote the C--B-bonded and C--N-bonded diamond(111)/\gls{cbn}(111) interfaces under $[111]$ tension, respectively, whereas C--B [001] denotes the C--B-bonded diamond(001)/\gls{cbn}(001) interface under $[001]$ tension. The stress--strain responses of C--B [111] are shown in \autoref{fig:main}(a), with the corresponding results for C--N [111] and C--B [001] provided in Supplementary Figures~S1 and S2, respectively. Fracture strength is defined as the peak tensile stress, and fracture strain as the corresponding strain. All systems initially exhibit an approximately linear elastic response, followed by nonlinear loading and a sharp stress decrease upon brittle fracture. Increasing diffusion systematically decreases both fracture strength and fracture strain (\autoref{fig:main}(b)--(c)). More importantly, the three systems exhibit distinct fracture locations, with failure occurring either at the interface or within the adjacent \gls{cbn} phase (\autoref{fig:main}(d)--(i)).

C--N [111] and C--B [001] both exhibit interface-controlled fracture across the investigated diffusion range, despite their markedly different tensile resistances. C--B [001] is the strongest of the three systems, with its fracture strength and fracture strain decreasing from 95.9~GPa and 15.49\% at 0.0\% diffusion to 35.6~GPa and 6.00\% at 50.0\% diffusion, respectively. Nevertheless, fracture remains localized within the interfacial region throughout the diffusion range (\autoref{fig:main}(d)--(e)). C--N [111], in contrast, exhibits substantially lower tensile resistance, with its fracture strength decreasing from 39.3 to 9.4~GPa and its fracture strain from 5.8 to 1.5\%. Its fracture likewise remains at or near the interface (\autoref{fig:main}(f)--(g)). Although both C--N [111] and C--B [001] fracture at the interface, their fracture strengths differ substantially. This indicates that the occurrence of interfacial fracture is determined not by the absolute strength of the interface alone, but by its resistance relative to that of the neighboring crystal planes.

C--B [111] exhibits a qualitatively different fracture behavior. Its fracture strength decreases from 50.3~GPa for the pristine interface to 15.8~GPa at 50.0\% diffusion, while the corresponding fracture strain decreases from 7.8 to 2.6\%. Unlike C--N [111] and C--B [001], however, the pristine C--B [111] heterostructure does not fracture at the interface. Instead, rupture occurs at a neighboring B--N plane within the \gls{cbn} phase, leaving the C--B-bonded interface intact (\autoref{fig:main}(h)). The fracture location is therefore transferred from the heterointerface into the adjacent bulk \gls{cbn}. This fracture mode changes progressively with increasing diffusion. As the initially sharp C--B interface becomes chemically intermixed, the preferred fracture region moves toward the interface.  At 50.0\% diffusion, fracture occurs within the intermixed interfacial region rather than at the neighboring B--N plane in \gls{cbn} (\autoref{fig:main}(i)). C--B [111] therefore exhibits a diffusion-induced transition from bulk fracture to interfacial fracture, in marked contrast to C--N [111] and C--B [001], which remain interface-controlled throughout the investigated diffusion range.

The diffusion-dependent trends and fracture-plane selection are retained under the additional loading conditions examined here, while the pristine structures show limited sensitivity to the model length along the loading direction. For C--B [111] and C--N [111], reducing the finite strain rate from $3\times10^{9}$ to $3\times10^{8}$~s$^{-1}$ or replacing finite-temperature \gls{md} with quasi-static loading changes the absolute fracture strength and strain but preserves their overall degradation with diffusion and their relative hierarchy (see \autoref{fig:rate}). More importantly, the fracture locations remain unchanged: pristine and weakly diffused C--B [111] fail within the neighboring \gls{cbn}, highly diffused C--B [111] fails within the intermixed interface, and C--N [111] remains interface-controlled. Likewise, doubling the model length along the loading direction from approximately 7 to 14~nm changes the fracture strengths of the ideal, diffusion-free systems by no more than 3.02~GPa and preserves both the strength hierarchy and fracture locations, including bulk fracture in C--B [111] and interfacial fracture in C--N [111] and C--B [001] (\autoref{fig:length}). The complete stress--strain responses at $3\times10^{8}$~s$^{-1}$ are provided for C--B [111] and C--N [111] in Supplementary Figures~S3 and S4, respectively, while their quasi-static responses are shown in Supplementary Figures~S5 and S6, respectively.

The key effect of diffusion in C--B [111] is therefore not only the reduction in fracture strength, but also the change in the selected fracture location. This bulk-to-interface transition indicates that the competition between the C--B-bonded interface and the neighboring \gls{cbn} plane evolves as interfacial mixing increases. The mechanical and energetic origins of this competition are examined below using atom-resolved stress fields and DFT separation energetics.

\subsection{Diffusion-driven transition from bulk to interfacial fracture}
\label{subsec:fracture-transition}

\begin{figure}[!t]
\centering
\includegraphics[width=\columnwidth]{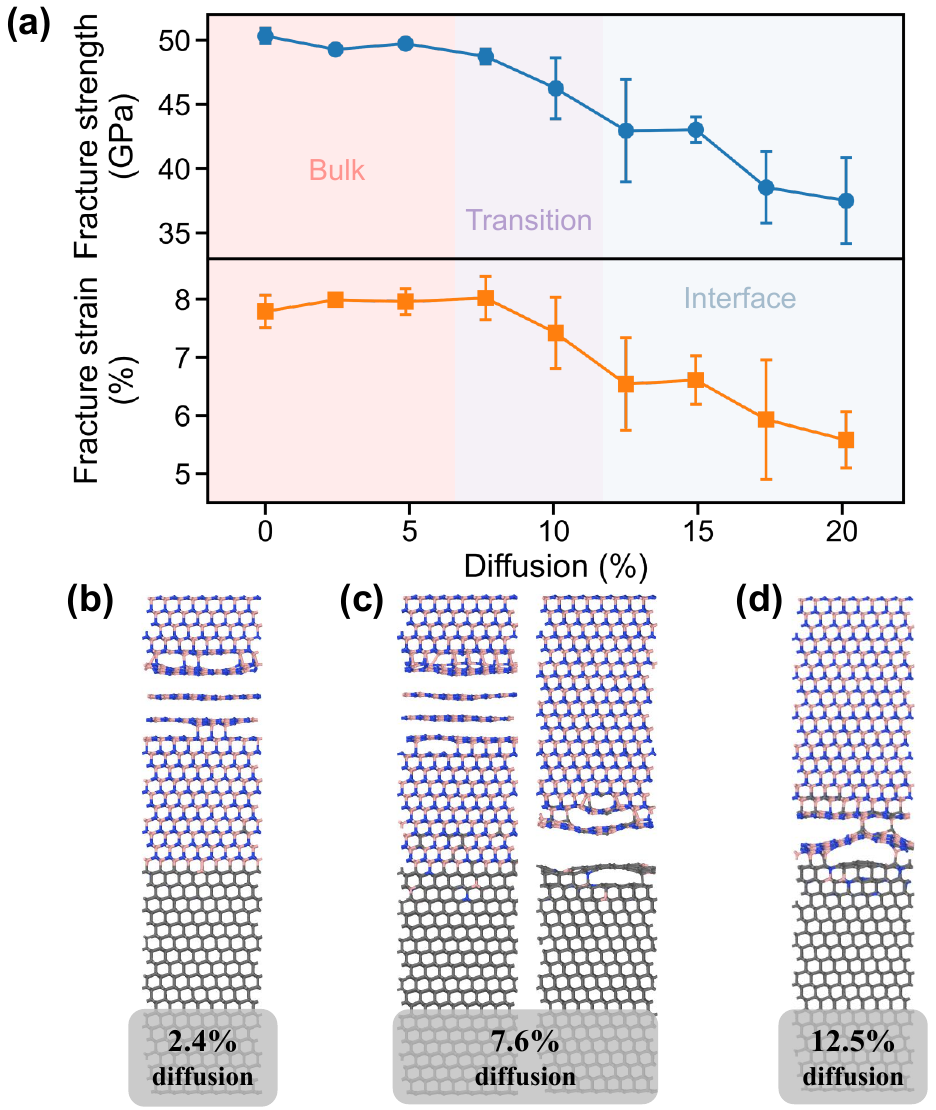}
\caption{Diffusion-driven transition from fracture within \gls{cbn} to interfacial fracture in C--B [111]. (a) Fracture strength and fracture strain as functions of diffusion percentage. Symbols and error bars denote means and standard deviations, respectively. The shaded regions indicate fracture within \gls{cbn} ($\leq4.9\%$), configuration-dependent fracture (7.6--10.1\%), and interfacial fracture ($\geq12.5\%$). (b--d) Representative fractured configurations at 2.4, 7.6, and 12.5\% diffusion. The two independently generated configurations at 7.6\% diffusion exhibit different fracture planes. Gray, pink, and blue spheres denote C, B, and N atoms, respectively.}
\label{fig:transition}
\end{figure}

The results in \autoref{fig:main} show that diffusion shifts the fracture location of C--B [111] from the adjacent \gls{cbn} phase toward the diamond/\gls{cbn} interface. To resolve this transition more precisely, additional simulations were performed using finer diffusion intervals in the low-diffusion range. As summarized in \autoref{fig:transition},  the configurations examined here exhibit three regimes: fracture within \gls{cbn} at low diffusion percentages, a configuration-dependent transition regime, and  interfacial fracture at higher diffusion percentages.

At diffusion percentages up to 4.9\%, all examined configurations fracture at a B--N plane inside the \gls{cbn} phase, while the nominal interface remains intact (\autoref{fig:transition}(b)). Within this regime, the fracture strength remains approximately 50~GPa and varies by less than 1~GPa, while the fracture strain remains within approximately 7.9\%. The absence of a systematic reduction indicates that low diffusion percentages do not weaken the interface sufficiently to change the controlling fracture plane. The C--B interface therefore continues to transmit tensile load into \gls{cbn}, where fracture is controlled by the neighboring B--N plane.

Between 7.6 and 10.1\% diffusion, the fracture plane becomes sensitive to the initial diffusion configuration. Independently generated structures at the same diffusion percentage can fracture either inside \gls{cbn} or within the mixed interfacial region. This competition is directly illustrated in \autoref{fig:transition}(c), where two configurations with 7.6\% diffusion select different fracture planes. At 7.6\% diffusion, configurations fracturing within \gls{cbn} have a mean fracture strength of approximately 48.9~GPa and a mean fracture strain of approximately 8.2\%, whereas those fracturing at the interface have corresponding values of approximately 48.4~GPa and 7.6\%, respectively. The occurrence of different fracture planes at the same diffusion percentage indicates that the diffusion percentage alone does not uniquely determine fracture location. These outcomes are consistent with competition between fracture within the mixed interface and at the neighboring B–N plane, suggesting that local atomic arrangements influence which region fails first. Similar configuration-sensitive pathway competition has been observed during the dislocation-to-ripplocation transformation in bilayer graphene~\cite{qiu2026metastability}.

At diffusion percentages of 12.5\% and above, fracture is consistently localized within the mixed interfacial region, as represented by the 12.5\% configuration in \autoref{fig:transition}(d). At this stage, diffusion has reduced the interfacial cohesion below that of the adjacent B--N plane, making the interface the preferred fracture region. Both fracture strength and fracture strain show an overall decrease with increasing diffusion, reaching approximately 37.5~GPa and 5.6\%, respectively, at 20.0\% diffusion. Relative to the pristine values of approximately 50.3~GPa and 7.8\%, these values correspond to reductions of about 25\% in strength and 28\% in fracture strain.

Taken together, the fracture properties exhibit a piecewise dependence on diffusion that closely follows the change in fracture location. Up to 4.86\% diffusion, the fracture strength and fracture strain form a plateau at approximately 49.8~GPa and 7.9\%, respectively. Within the transition regime of 7.6--10.1\%, their mean values begin to decrease, but the reduction remains moderate and is accompanied by increased configuration-to-configuration scatter. At 10.1\% diffusion, the strength and strain are approximately 46.3~GPa and 7.4\%, corresponding to reductions of approximately 8\% and 5\%, respectively, relative to the pristine system. Once fracture becomes interface-controlled at higher diffusion percentages, both quantities decline more substantially, reaching approximately 37.5~GPa and 5.6\% at 20.0\% diffusion. These values represent reductions of about 25\% in fracture strength and 28\% in fracture strain relative to the pristine interface, showing that most of the mechanical degradation develops after the mixed interface becomes the controlling fracture region. The evolution therefore reflects not only progressive interfacial weakening, but also a change in the mechanically controlling plane from the neighboring B--N plane to the interface. The complete stress--strain curves,  corresponding atom-resolved tensile-stress distributions and atomic fractured configurations are provided in Supplementary Figures~S7, S8 and S9, respectively.

\subsection{Atom-resolved stress and energetic competition}
\label{subsec:stress-energy}

\begin{figure}[!t]
\centering
\includegraphics[width=\columnwidth]{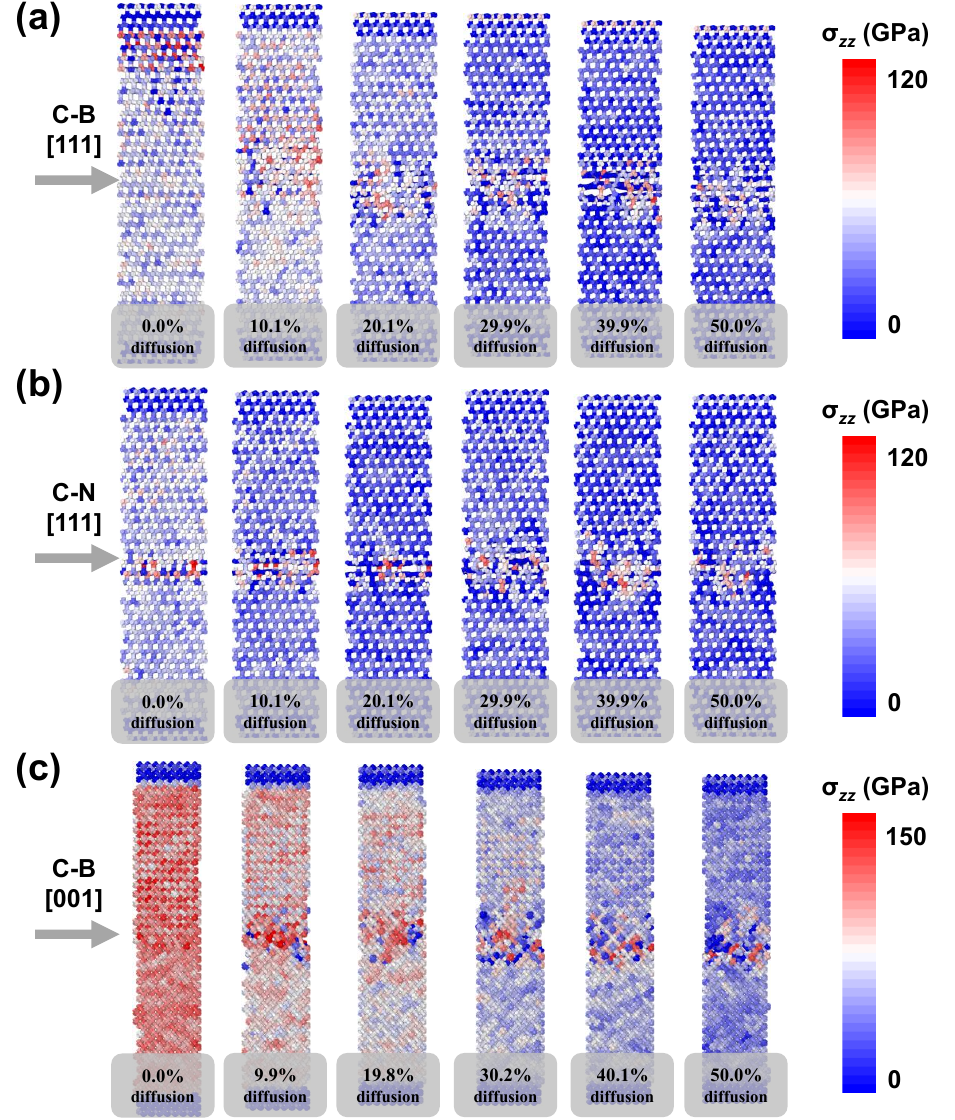}
\caption{Representative atom-resolved tensile-stress fields $\sigma_{zz}$ in the last intact frame immediately before fracture for (a) C--B [111], (b) C--N [111], and (c) C--B [001], with diffusion percentage increasing from left to right. Gray arrows mark the nominal diamond/\gls{cbn} interfaces. Panels (a,b) use a stress range of 0--120~GPa, whereas panel (c) uses 0--150~GPa.}
\label{fig:stressfield}
\end{figure}

Atom-resolved tensile stress fields track how interfacial bonding, crystallographic orientation, and diffusion-induced intermixing redistribute the applied load and shift the stress concentration preceding fracture. \autoref{fig:stressfield} shows the atomic tensile-stress component $\sigma_{zz}$ in the last intact frame immediately before the macroscopic stress drop for C--B [111], C--N [111], and C--B [001].

In pristine C--B [111], the highest tensile stresses are concentrated on a B--N plane within the adjacent \gls{cbn} region rather than at the nominal diamond/\gls{cbn} interface (see 0.0\% diffusion result in \autoref{fig:stressfield}(a)). This highly stressed plane is subsequently selected as the fracture plane (\autoref{fig:main}(h)), showing that the intact C--B-bonded interface transfers the applied load into the adjacent \gls{cbn}. At low diffusion percentages, the dominant stress concentration remains on the \gls{cbn} side. As the diffusion percentage increases, the stress field becomes more heterogeneous and its dominant concentration progressively shifts toward the chemically mixed interface boundary. At 50.0\% diffusion, highest tensile stresses are localized within the intermixed region, consistent with the interfacial fracture observed in \autoref{fig:main}(i).

By contrast, C--N [111] exhibits interface-centered stress localization throughout the diffusion range (\autoref{fig:stressfield}(b)). Even in the pristine structure, the highest $\sigma_{zz}$ values occur at or immediately adjacent to the C--N-bonded interface. Increasing diffusion broadens the stressed region and enhances its atomic-scale heterogeneity, but does not shift the dominant stress concentration into the interior of \gls{cbn}, consistent with the persistence of interfacial fracture. C--B [001] likewise retains interface-centered stress localization at all investigated diffusion percentages (\autoref{fig:stressfield}(c)) but sustains substantially higher atom-resolved tensile stresses immediately before fracture than the two systems loaded along [111]. Accordingly, an upper limit of approximately 150~GPa is reached for C--B [001], compared with 120~GPa for C--B [111] and C--N [111]. This higher local stress scale is consistent with the considerably greater macroscopic fracture strength of C--B [001]. Taken together, the stress fields distinguish the  change in the location of highly stressed regions in C--B [111] from the persistent interface-centered stress localization in C--N [111] and C--B [001]. Because the maps represent the last intact frame of each trajectory, they correspond to different macroscopic stress and strain levels. They therefore provide qualitative evidence of local stress distributions immediately before failure, consistent with the observed fracture locations.

\begin{figure}[!t]
\centering
\includegraphics[width=\columnwidth]{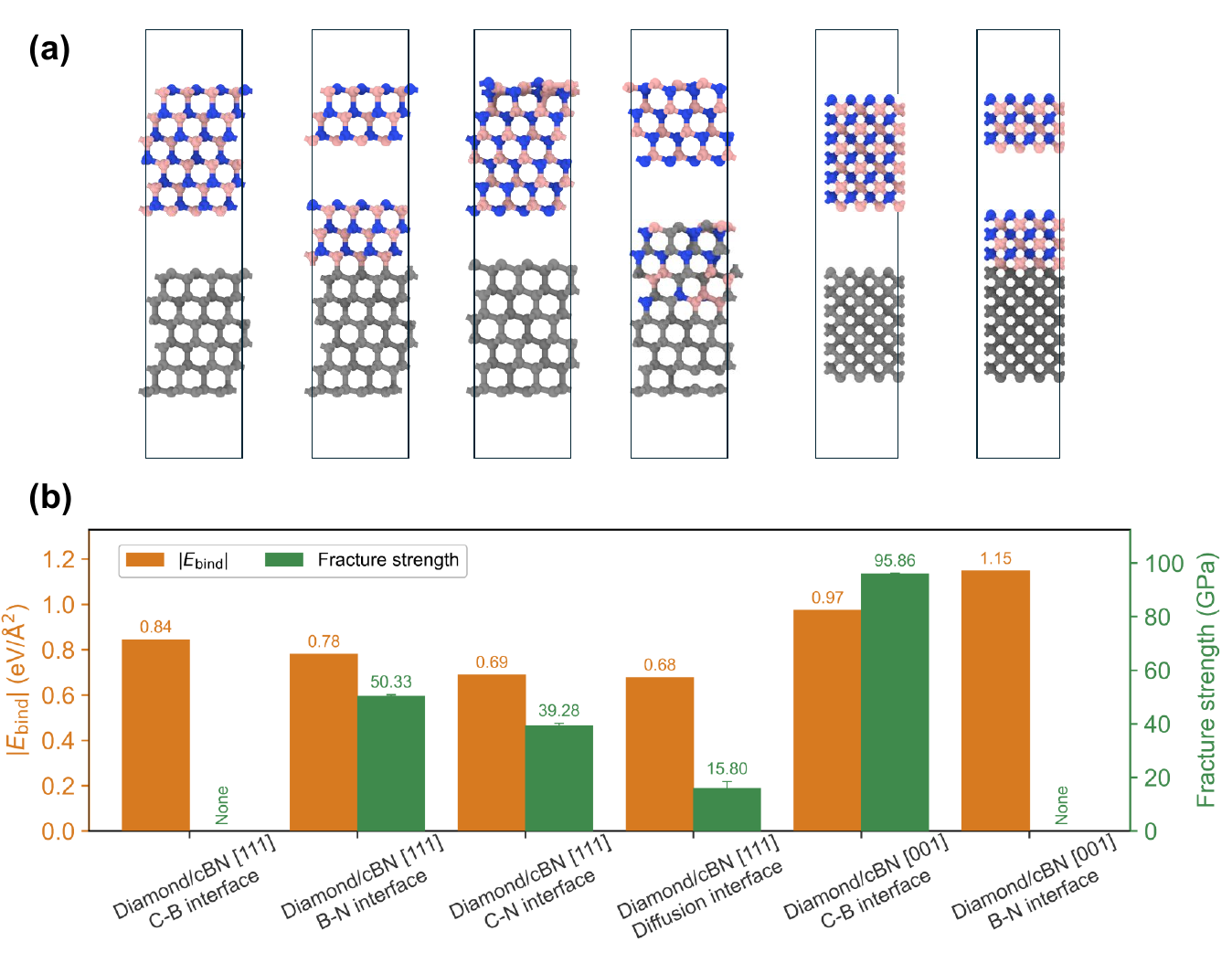}
\caption{Binding-energy analysis of competing separation planes in diamond/\gls{cbn} heterostructures. (a) DFT-relaxed separated configurations for, from left to right, the pristine C--B [111] interface, the neighboring B--N plane in C--B [111], the pristine C--N [111] interface, the 50.0\%-diffused C--B [111] interface, the pristine C--B [001] interface, and the neighboring B--N plane in C--B [001]. (b) DFT binding-energy magnitudes $\lvert E_{\mathrm{bind}}\rvert$ (orange, left axis) and the MD fracture strengths associated with the observed separation planes (green, right axis). A larger binding-energy magnitude indicates greater resistance to separation. No fracture strength is assigned when the corresponding plane is not selected during fracture. }
\label{fig:binding}
\end{figure}

The \gls{dft} binding energies in \autoref{fig:binding} provide an energetic measure of the competition among the candidate separation planes. Because the binding energies are negative under the adopted definition, a larger magnitude, $\lvert E_{\mathrm{bind}}\rvert$, represents stronger cohesion. Within each heterostructure, the plane with the smaller binding-energy magnitude is therefore expected to separate more readily. The binding energies are used here to rank the competing fracture planes rather than as a one-to-one predictor of fracture strength.

\Gls{dft} binding-energy calculations establish the energetic hierarchy between the competing separation planes within the diamond/\gls{cbn} heterostructures (\autoref{fig:binding}). To correspond directly to the fracture simulations, the comparison is restricted to the interfaces and their neighboring B--N planes. For pristine C--B [111], the C--B interface has a binding-energy magnitude of 0.84~eV~\AA$^{-2}$, whereas the neighboring B--N plane has a smaller value of 0.78~eV~\AA$^{-2}$ (\autoref{fig:binding}(a,b)). The latter is therefore energetically easier to separate, consistent with fracture occurring inside \gls{cbn} at a strength of 50.3~GPa. By comparison, the pristine C--N [111] interface has a lower binding-energy magnitude of 0.69~eV~\AA$^{-2}$ and fractures directly at the interface at 39.3~GPa. At 50.0\% diffusion, the binding-energy magnitude of the C--B [111] interface decreases from 0.84 to 0.68~eV~\AA$^{-2}$, making the chemically mixed interface the preferred separation region and reducing the fracture strength to 15.8~GPa.

The energetic ordering is reversed in C--B [001]. Its C--B interface and neighboring B--N plane have binding-energy magnitudes of 0.97 and 1.15~eV~\AA$^{-2}$, respectively. The interface is therefore approximately 16\% weaker than the adjacent B--N plane and is selected as the fracture plane, although its fracture strength remains as high as 95.9~GPa. These comparisons show that, within each heterostructure, the plane with the smaller binding-energy magnitude corresponds to the preferred fracture location. However, the binding energy should not be interpreted as a universal predictor of the absolute fracture strength. For example, the pristine C--N [111] interface and the 50.0\%-diffused C--B [111] interface have similar binding-energy magnitudes of 0.69 and 0.68~eV~\AA$^{-2}$, but their fracture strengths differ substantially. This difference reflects the disordered bond topology and broader stress redistribution introduced by diffusion.

According to \autoref{fig:binding}, the relative energetic ordering in ideal heterointerfaces can be summarized by a simple cohesion ratio,
\begin{equation}
\eta_E=
\frac{\left|E_{\mathrm{bind}}^{c\mathrm{BN}}\right|}
{\left|E_{\mathrm{bind}}^{\mathrm{int}}\right|},
\end{equation}
where $E_{\mathrm{bind}}^{c\mathrm{BN}}$ and $E_{\mathrm{bind}}^{\mathrm{int}}$ denote the binding energies of the relevant neighboring B--N plane and the interface, respectively. When $\eta_E>1$, the interface has a smaller binding-energy magnitude than the neighboring B–N plane;  when $\eta_E<1$, the neighboring B–N plane has the smaller magnitude. For the pristine C–B interfaces examined here, these energetic orderings are consistent with interfacial fracture in C–B [001] and fracture within adjacent \gls{cbn} in C–B [111], respectively. The limiting case $\eta_E\approx1$ indicates similar separation energies for the two planes, but does not necessarily imply equal peak tensile strengths.  This ratio provides an empirical energetic indicator of fracture-plane selection for the examined structures, but is not intended to predict the absolute fracture strength.

\subsection{Bonding origin of fracture selection}
\label{subsec:electronic}

\begin{figure}[!t]
\centering
\includegraphics[width=\columnwidth]{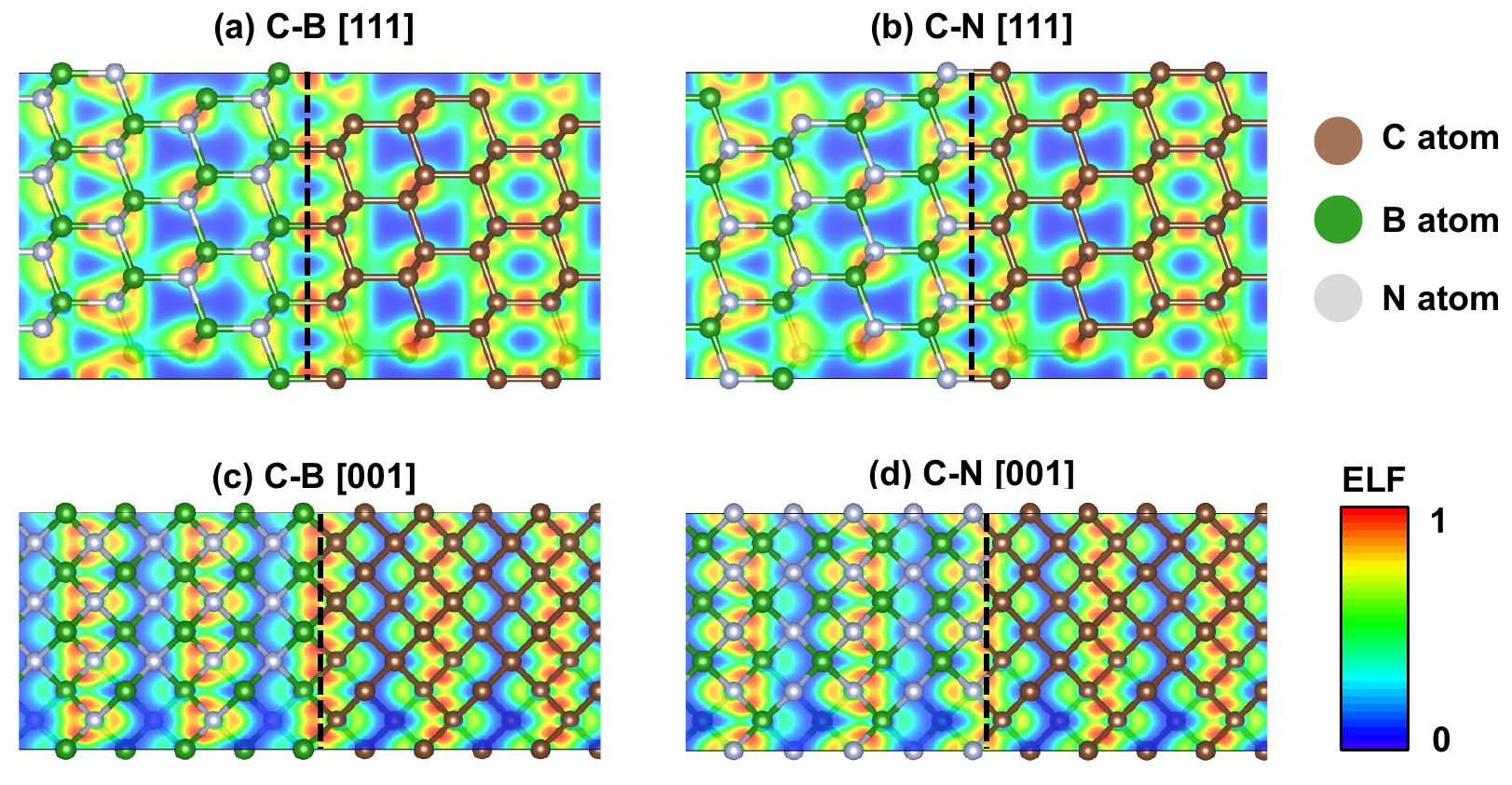}
\caption{\gls{elf} maps of pristine diamond/\gls{cbn} interfaces: (a) C--B [111], (b) C--N [111], (c) C--B [001], and (d) C--N [001]. Brown, green, and light-gray spheres represent C, B, and N atoms, respectively. All panels use the same ELF scale from 0 (low localization, blue) to 1 (high localization, red). The black dashed lines mark the location of the interfaces.}
\label{fig:elf}
\end{figure}

The \gls{elf} maps in \autoref{fig:elf} show how interfacial termination and crystallographic orientation alter the spatial distribution of localized valence electrons. Both C--B [111] and C--N [111] exhibit regions of elevated ELF along their cross-interface bonds, indicating covalent coupling between diamond and \gls{cbn}. In C--B [111], these regions form comparatively continuous bridges across the interface (\autoref{fig:elf}(a)). This continuity is qualitatively consistent with the strong cohesion of the pristine C–B [111] interface indicated by the binding-energy calculations. C--N [111], by contrast, exhibits a more asymmetric and less continuous localization pattern across the interface (\autoref{fig:elf}(b)). Both C--B and C--N are heteropolar covalent bonds, but they differ in electronic filling, orbital contributions, and interfacial charge redistribution. Previous first-principles studies found predominantly $p$-orbital coupling and stronger covalent character for C--B interfaces, whereas C--N interfaces contain additional N-$2s$ contributions; C--B terminations consequently exhibit higher formation and adhesion energies than C--N terminations~\cite{chen2015misfit,zhao2019electronic,milne2025interface}. These differences are consistent with the present binding-energy hierarchy and the contrasting fracture locations of the two (111) interfaces.

The (001)-oriented maps further demonstrate that termination cannot be interpreted independently of interface geometry. Viewed in the present projection, the (001) interfaces exhibit a denser, zigzag-like network of inclined cross-interface bonds than the (111) interfaces. Counting the bonds crossing a unit interface area in the ideal coherent structures gives areal bond densities of approximately $0.31$ and $0.18$~bonds~\AA$^{-2}$ for the (001) and (111) interfaces, respectively. In C--B [001], regions of elevated ELF follow this dense, connected bond network across the interface (\autoref{fig:elf}(c)), consistent with its high binding-energy magnitude and fracture strength. C--N [001] adopts the same zigzag-like (001) bond topology but retains a more asymmetric termination-dependent localization pattern (\autoref{fig:elf}(d)). C--N [001] is included here only as an electronic-structure comparison. The \gls{elf} results provide a qualitative picture of termination- and orientation-dependent bonding in the pristine interfaces, consistent with the calculated binding-energy trends. 

\section{Conclusions}
\label{sec:conclusions}

In this work, a machine-learned interatomic potential was developed by fine-tuning an atomistic foundation model and assessed against independent \gls{dft} tensile responses, including a (001)-oriented interface absent from the task-specific fine-tuning dataset. The potential enabled systematic tensile-fracture simulations of diamond/\gls{cbn} heterostructures across different interfacial bonding configurations, crystallographic orientations, and diffusion percentages.

The results highlight competition between the interface and neighboring crystal planes in determining fracture location. In pristine C--B [111], fracture initiates at a neighboring B--N plane within \gls{cbn}, leaving the C--B-bonded interface intact. As diffusion increases from 0.0 to 50.0\%, the fracture strength decreases nonlinearly from 50.3 to 15.8~GPa and the fracture strain decreases from 7.8 to 2.6\%. Closer sampling reveals three distinct regimes in this bulk-to-interfacial fracture transition. Fracture remains within \gls{cbn}, with nearly unchanged strength and strain, up to 4.9\% diffusion. Between 7.6 and 10.1\%, either the neighboring B--N plane or the interface can fracture depending on the initial diffusion configuration, whereas fracture occurs within the interfacial region at the sampled diffusion percentages of 12.5\% and above.  In contrast, C--N [111] is mechanically weaker and remains interface-controlled throughout the investigated diffusion range. C--B [001] also fractures within the interfacial region, but exhibits substantially greater fracture strength and strain than the two (111)-oriented systems. Thus, interfacial fracture does not necessarily imply low tensile strength.

Atom-resolved stress fields show a shift in the location of highly stressed regions from the \gls{cbn} region toward the mixed interface with increasing diffusion in C--B [111], whereas C--N [111] and C--B [001] retain interface-centered stress localization. \Gls{dft} binding energies provide a complementary energetic interpretation: the pristine C--B [111] interface is more strongly bound than the neighboring B--N plane, while the lower interfacial binding-energy magnitude calculated at 50.0\% diffusion is consistent with the observed interfacial fracture. \Gls{elf}  distributions further illustrate termination- and orientation-dependent electron localization in the pristine interfaces, providing a qualitative bonding interpretation consistent with the binding-energy results.

For C--B [111] and C--N [111], the diffusion-dependent strength ordering and fracture locations are retained under the finite strain rates and quasi-static loading conditions examined here. Increasing the model length along the loading direction from approximately 7 to 14~nm likewise produces only minor changes in the fracture strength of the pristine  structures and does not alter the selected fracture plane. Overall, these results clarify the roles of interfacial termination, diffusion, and crystallographic orientation in intrinsic tensile failure and provide an atomistic basis for understanding fracture-plane selection in coherent diamond/\gls{cbn} heterostructures.


\section*{Acknowledgments}
The authors acknowledge the financial support from the National
Natural Science Foundation of China (12325204). W. Qiu acknowledges the postdoctoral fellowships of the Sackler Center for Computational Molecular and Materials Science and the Ratner Center for Single Molecule Science at Tel Aviv University. F. Xie acknowledges President's Fund Project of China Academy of Engineering Physics (YZJJZQ2023013) and National Natural Science Foundation of China-Ye QiSun Science Foundation (U2441249). We acknowledge the Open Source Supercomputing Center of S-A-I for providing the computing resources used in this work.


\section*{Declaration of competing interest}

The authors declare that they have no known competing financial
interests or personal relationships that could have appeared to
influence the work reported in this paper.



\balance
\bibliographystyle{elsarticle-num}

\end{document}